\documentstyle[prl,manuscript,floats,aps,epsf]{revtex}

\newcommand{\N}{{\mathchoice{\hbox{$\sf\textstyle I\hspace{-.15em}N$}}     
{\hbox{$\sf\textstyle I\hspace{-.15em}N$}}
{\hbox{$\sf\scriptstyle I\hspace{-.10em}N$}}
{\hbox{$\sf\scriptscriptstyle I\hspace{-.11em}N$}}}}
\newcommand{\R}{{\mathchoice{\hbox{$\sf\textstyle I\hspace{-.15em}R$}}     
{\hbox{$\sf\textstyle I\hspace{-.15em}R$}}
{\hbox{$\sf\scriptstyle I\hspace{-.10em}R$}}
{\hbox{$\sf\scriptscriptstyle I\hspace{-.11em}R$}}}}

\begin{document}
\draft
\title{A black hole in two-dimensional space-time}
\author{F.~Vendrell$^\ast$}
\address{Blackett Laboratory, Imperial College, Prince Consort Road,
London SW7 2BZ, U.K.}
\date{May 19, 1997}
\maketitle

\begin{abstract}
An imploding shell of radiation is shown to create a 2-D black hole
within the framework of the ``$R=T$'' theory.
The radius of the horizon is given by $r_H=\frac{1}{2M}$, where $M$ is the 
mass of the black hole.
The topology of the central singularity is that of a corner.
The radiation emitted very far from the black hole is thermal with temperature
$\Theta^{\rm rad}=\frac{\hbar M}{2\pi k}$.
The back-reaction problem is solved to one-loop order.
\end{abstract}



Black holes in 2-D space-times are toy models to understand the quantum 
physics of 4-D black holes, in particular their thermal radiation \cite{Ha}.
Since the general theory of relativity has no physical contents in two 
dimensions \cite{Co}, other 2-D gravity theories have been considered, in 
particular string theory for which there exist 2-D black hole 
solutions \cite{CGHS}.
It is possible, however, to extract from the equations of general 
relativity a gravity theory by considering the formal limit 
$D\rightarrow 2$, where $D$ is the dimension of 
space-time \cite{Ma}.
One obtains the equation
\begin{eqnarray}
R(x) = 8\pi G\,T^{\rm cl}(x),
\label{RT}
\end{eqnarray}
where $c=1$, $G$ is Newton's constant and where
$T^{\rm cl}(x)$ is the trace of the classical energy-momentum tensor 
$T_{\mu\nu}^{\rm cl}(x)$.
This equation is supplemented by the continuity equation
\begin{eqnarray}
\nabla^\mu T_{\mu\nu}^{\rm cl} (x) = 0,
\label{CE}
\end{eqnarray}
which follows also from the equations of general relativity.

A black-hole solution was obtained from these equations by Mann et al.~by 
considering a static distribution of matter localised in space \cite{MST}.
This black hole is an extension and generalisation of the solution of
Brown et al.~\cite{BHT}.
It has a number of similarities with the Schwarzschild
black hole, in particular, it may be created from a collapse of matter.
However, there is no {\it global} set of 
coordinates for which the metric is Minkowskian very far from the black hole,
and, furthermore, its thermal radiation does not originate from a 
{\it global} vacuum \cite{ST}.
At least in this sense this black hole differs from the 4-D case.

There is, however, a 2-D black hole based on Eqs.~(\ref{RT}) and (\ref{CE}) 
which satisfies the properties above, as shown in the present Letter.
This is formed from an imploding shell of radiation symmetric under 
space-reflection.
The metric obtained is discontinuous when the shell is infinitesimally 
thin, as in the 3-D circularly collapse \cite{3D},
but contrary to the 4-D spherically collapse where it is possible to require 
its continuity (but not the continuity of its derivative) \cite{Sy}.
The 2-D metric exhibits a coordinate singularity, where the horizon is located,
and a central {\it real} singularity as well.
The interior and exterior of the black hole are causally disconnected regions.
A massless or massive particle inside the black hole will inextricably
fall into the central singularity within a finite affine parameter or time.
The topology of the singularity is that of a corner.

I also show that this 2-D black hole emits a thermal 
radiation whose temperature is proportional to its mass.
The back-reaction on space-time may easily be analysed to one-loop order
due to the simplicity of the model.


Consider a 2-D space-time which is symmetric under space-reflection 
with respect to an origin and which is covered {\it globally} by the set of
coordinates $(t,r)$, where $t\in\R$ is the time and $r\geq 0$ is the distance 
to the origin.
I assume that this space-time may also be covered by a set 
of {\it conformal} coordinates  $x^\pm=x^0\pm x^1\in\R$:
\begin{eqnarray}
ds^2= C(x)\,dx^+\,dx^-,
\label{conformal}
\end{eqnarray}
such that, if $x^+\leq x^+_0$ and $x^1\geq0$,
\begin{eqnarray}
x^\pm = t \pm r,
\label{rtx<}
\end{eqnarray}
where $x^+_0>0$ is an arbitrary constant.

Consider now a imploding shell of radiation localised at $x^+=x^+_0$ 
and defined in accordance with Eq.~(\ref{CE}) by
\begin{eqnarray}
T^{\rm cl} (x) =\frac{M}{2\pi G}\,\delta(x^+-x^+_0),
\label{SW}
\end{eqnarray}
where $M>0$ is a constant.
Equation (\ref{RT}) implies that the curvature is infinite where the shell is 
localised and that the conformal factor $C(x)$ satisfies the differential 
equation 
\begin{eqnarray}
\partial_+\partial_- \ln \vert C(x)\vert = M\,C(x)\,\delta(x^+-x^+_0).
\label{C(x)}
\end{eqnarray}
I also assume that the space-time is Minkowskian inside the imploding shell:
\begin{eqnarray}
C(x)= 1, &\hspace{10mm}&\mbox{if $x^+<x^+_0$.}
\label{BC} 
\end{eqnarray}
The problem consists of obtaining the conformal factor $C(x)$ outside
the imploding shell and then extending there the $(t,r)$ coordinates
in a sensible way.
Then the entire accessible space-time will be known, since this is
represented in the plane $(x^+,x^-)$ by the set of points whose radius
$r$ is larger than zero, also called the {\it physical region}.
Equation (\ref{rtx<}) implies that the region $x^1<0$ is unphysical if 
$x^+\leq x^+_0$.

The problem given by Eqs.~(\ref{C(x)}) and (\ref{BC}) is not well defined
for {\it every} definition of the delta function (or structure of the imploding
shell), and if it is, the solution will depend on that definition (see below).
I assume here that the delta function is defined by
\begin{eqnarray}
\int_{-\infty}^{+\infty}du \, f(u)\,\delta(u)= 
\lim_{\epsilon\rightarrow 0^+} f(\epsilon) \equiv f(0^+),
\end{eqnarray}
where the test function $f$ is continuous except maybe at $u=0$.
This delta function may be represented by the limit of a series of normalised 
functions $\{\delta_n\}_{n\in\N}$ vanishing for negative arguments 
(see fig.~1 for a particular representation).
\begin{figure}
\centerline{
\epsfysize=2.4truein
\epsfbox{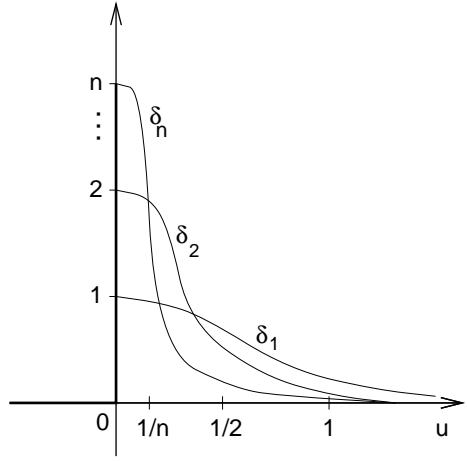}}
\caption{The normalised functions $\delta_n$, $n\in\N$.}
\end{figure}
Define also the theta function $\theta$ by
\begin{eqnarray}
\theta(u) &=& \left\{ 
\begin{array}{rcl}
0, & \hspace{5mm} &\mbox{if $u< 0$,} \\ [2mm]
1, & &\mbox{if $u>0$,}
\end{array} \right.
\end{eqnarray}
and the function $m(x^+) = M\,\theta(x^+-x^+_0)$.
Then integrating Eq.~(\ref{C(x)}) once gives
\begin{eqnarray}
\partial_- \ln \vert C(x)\vert = 
m(x^+)\,\left.C(x)\right\vert_{x^+=x^+_0+0^+},
\end{eqnarray}
where the boundary condition (\ref{BC}) has been used.
This equation would have been different for another definition of the 
delta function \cite{DF}.
Integrating again gives the discontinuous line element
\begin{eqnarray}
ds^2= \frac{dx^+\,dx^-}{1-m(x^+)\,(x^--\Delta)},
\label{dsx}
\end{eqnarray}
where $\Delta$ is a constant.

This result was first obtained in Ref.~\cite{Ve} but is valid in our case 
only in the physical region which will be fixed below.
The value of the conformal factor at $x^+=x^+_0$ may not be determined.
This is because the solution of Eqs.~(\ref{C(x)}) and (\ref{BC}),
when the delta function is replaced by one of the functions $\delta_n$,
depends strongly on $x$ in the interval $[x^+_0,x^+_0+1/n]$ if $n\gg1$.
The solution becomes discontinuous in the limit $n\rightarrow \infty$ only.
The constant $\Delta$ is arbitrary since the form of both the metric 
(\ref{conformal}) and the energy-momentum tensor (\ref{SW}) 
are invariant under translation of $x^-$.
The conformal factor $C(x)$ is singular at 
$x^-= x^-_H\equiv \frac{1}{M}+\Delta$, if $x^+> x^+_0$, and is negative 
only if $x^+> x^+_0$ and $x^->x^-_H$.
Furthermore, it is equal to one on the surface defined by 
$x^-=\Delta$ and $x^1\geq 0$.
This implies that the energy of the shock wave (\ref{SW}) flowing through
this surface is given by
\begin{eqnarray}
E = 2 \int_{\Delta}^\infty dx^+ \,
\left. T^{{\rm cl}}(x)\right\vert_{x^-=\Delta}
= \frac{M}{\pi G}\cdot
\label{EM}
\end{eqnarray}


In order to understand the structure of the space-time obtained better, the 
two sets of coordinates $y_I$ and $y_E$ are defined for $x^->x_H^-$ and 
$x^-<x_H^-$ respectively by
\begin{eqnarray}
&\mbox{\hspace{3mm}}&
\left\{
\begin{array}{rcl}
y^+_{^{^I_E}}(x^+) &=& x^+, \\ [2mm]
y^-_{^{^I_E}}(x^-) &=& \pm\frac{1}{M}\,\ln \left\vert\,x^-_H-x^- \,\right\vert,
\end{array} \right. 
\label{y}
\end{eqnarray}
where $y_{I,E}^\pm\in\R$ and the upper (lower) sign is related to 
the upper half-plane $x^->x_H^-$ (lower half-plane $x^-<x_H^-$)
[the subscripts are related to the Interior and Exterior of the black hole]. 
In these coordinates the line element (\ref{dsx}) becomes, if $y^+> x^+_0$,
\begin{eqnarray}
ds^2 = \left\{ 
\begin{array}{ccl}
-dy_I^+\,dy_I^-, & \mbox{\hspace{3mm}}& \mbox{if $x^->x_H^-$,} \\[3mm]
+dy_E^+\,dy_E^-, & &\mbox{if $x^-<x_H^-$.}
\end{array}
\right.
\label{dsy}
\end{eqnarray}
This implies that the geodesics expressed in terms of the $x$ and $y_{I,E}$ 
coordinates are straight lines in the half-planes $x^+<x^+_0$ and $x^+> x^+_0$ 
respectively.
The geodesics are continuous functions of their affine parameter, but
the slope of non-null geodesics are discontinuous across the shell because
the curvature is infinite there.
Their velocity $\dot{x}_f$ just outside the shell is given in terms of their 
initial velocity $\dot{x}_i$ by $\dot{x}_f^-=\dot{x}_i^-$ and
\begin{eqnarray}
\dot{x}_f^+= M\,\dot{x}_i^+\, \left(\,x^-_H-x^-_c\,\right),
\end{eqnarray}
if $x_c$ is where they cross the shell.

It takes an {\it infinite} proper-time or affine parameter for
a time-like or null geodesic to reach or to move away from the coordinate
singularity at $x^-=x^-_H$.
This implies that the two regions $x^->x^-_H$ and $x^-<x^-_H$
are {\it causally disconnected} if $x^+> x^+_0$.
There is no set of local coordinates centred at $x^-=x^-_H$ for which the 
metric is not singular there.
Although space-time is flat everywhere if $x^+>x^+_0$, it is not possible 
to find a set of {\it global} coordinates for which {\it i)} the metric is 
Minkowskian {\it} and {\it ii)} which cover the half-plane $x^+>x^+_0$.
At least two such set of coordinates must be used, as for instance 
the $y_{I,E}$ coordinates.

The $(t,r)$ coordinates are now extended in the right half-plane $x^+> x^+_0$.
Contrary to the higher-dimensional cases, there is no differential equation 
for $r=r(x^+,x^-)$ which follows from the equations of motion.
This stems from the fact that, in two dimensions, there is no angular
contribution to the metric.
In consequence, the function $r(x^+,x^-)$ is left undetermined 
for $x^+> x^+_0$.
This is also the case for the function $t(x^+,x^-)$.
It is however natural to assume that the curves $r=const.$ are geodesics 
outside the imploding shell as well. 
By extending them across the shell continuously, the spatial coordinate $y^1$ 
is related to the radius by
\begin{eqnarray}
y^1_{^{^I_E}}(r) = \frac{x^+_0}{2}\mp 
\frac{1}{2M}\log\left\vert\,2r+x^-_H-x^+_0\,\right\vert.
\end{eqnarray}
I require this expression to take the more familiar form
\begin{eqnarray}
y^1_{^{^I_E}}(r) = r_0 \mp r_H \log\left\vert\,r-r_H\,\right\vert,
\label{rt}
\end{eqnarray}
where $r_0=\displaystyle x^+_0/2\mp\ln 2/(2M)$, so that the constants $x_H^-$ 
and $r_H$ may be fixed:
\begin{eqnarray}
\begin{array}{rclcrcl}
x_H^- &=& \displaystyle x^+_0 - \frac{1}{M}, &\hspace{5mm}&
r_H &=& \displaystyle \frac{1}{2M}\cdot
\label{rH}
\end{array}
\end{eqnarray}
The time is defined by $t = \mp y^0_{^{^I_E}}$ if $x^+>x^+_0$.
The coordinates $(x^+,x^-)$ and $(t,r)$ are thus related through Eqs.~(\ref{y})
and (\ref{rt}) by
\begin{eqnarray}
e^{-2Mt} &=& e^{\pm M x^+} \left\vert\, x^-_H-x^-\,\right\vert,
\label{rtx>*}\\ [2mm]
2r-\frac{1}{M}&=& 
e^{\mp M (x^+-x^+_0)}\left(\,x^-_H-x^-\,\right).
\label{rtx>}
\end{eqnarray}
These equations show that the two curves $r=r_H$ and $x^-=x^-_H$ coincide if
$x^+>x^+_0$ and that $x^-=x^-_H$ implies $t=+\infty$ if $x^+>x^+_0$.
The amount of time $\delta t(r)$ for a curve $r=const.$ to cross the shell 
is given from Eqs.~(\ref{rtx<}), (\ref{rtx>*}) and (\ref{rtx>}) by
\begin{eqnarray}
\delta t(r)= r-\frac{1}{2M}\ln\left\vert\, r-\frac{1}{2M}\,\right\vert
\mp r_0 -x^+_0,
\end{eqnarray}
and diverges when $r$ tends to $r_H$.
The accessible space-time and its topology are now completely specified.
They are represented in fig.~2.

\begin{figure}
\centerline{
\epsfysize=3.375truein
\epsfbox{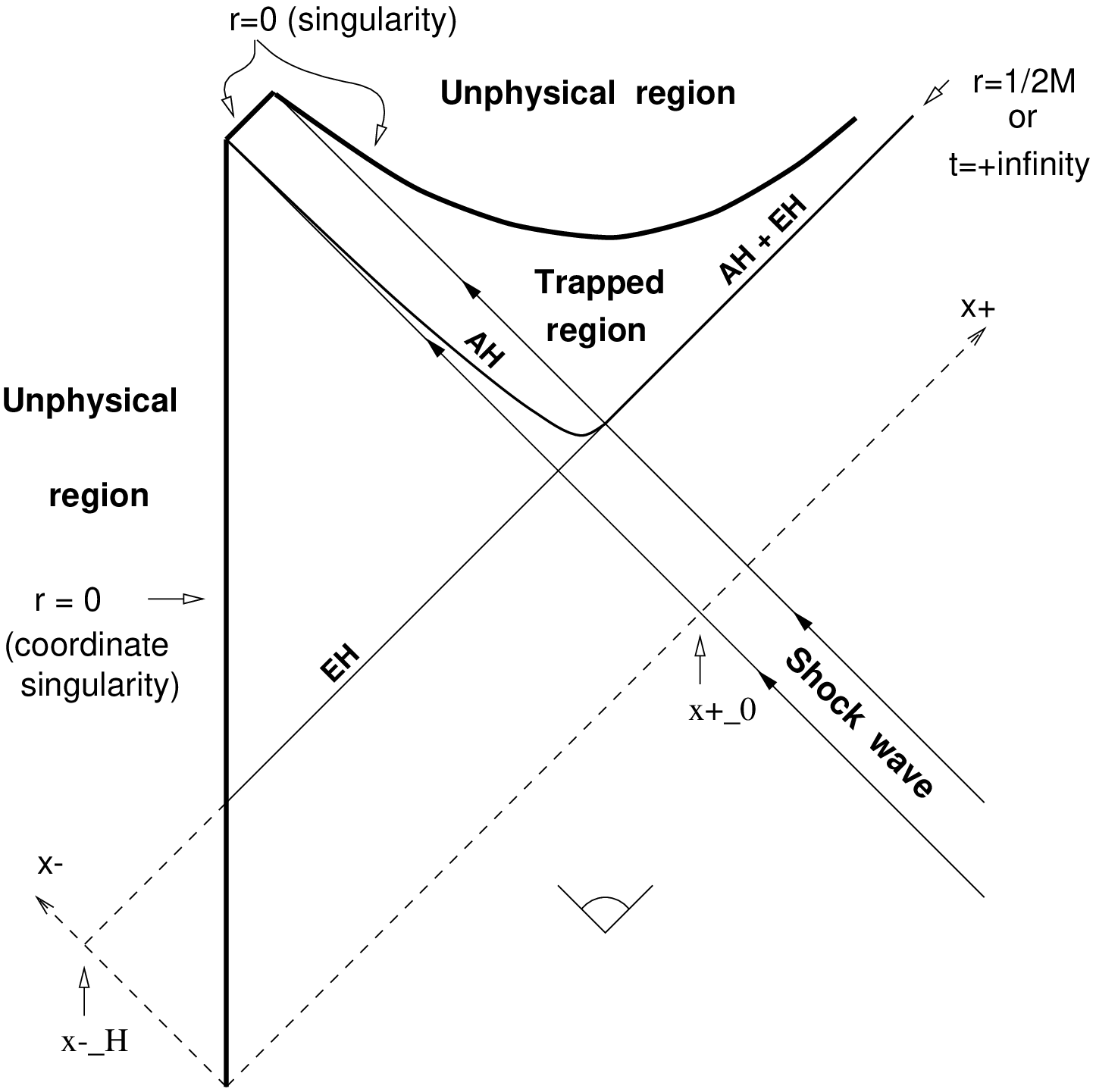}}
\vspace{5mm}
\caption{Space-time diagram of the 2-D black hole.
The light cones are looking upwards everywhere.
The curves denoted by AH and EH are the apparent and event horizons 
respectively.}
\end{figure}

Inside the collapsing shell the line element in the $(t,r)$ coordinates
is Minkowskian, and outside of it is given from Eqs.~(\ref{dsy}) and 
(\ref{rt}) by
\begin{eqnarray}
ds^2 = {\rm sgn}\left(r-\frac{1}{2M}\right)\, 
\left[\,dt^2- \frac{dr^2}{\left(\,2Mr-1\,\right)^2}\,\right].
\label{dsrt}
\end{eqnarray}
This line element is singular at $r=r_H$.
The region $r\leq r_H$ outside the imploding shell is a 
{\it trapped} region because all ingoing and outgoing null geodesics with 
initial point in this region will be contained in it, and so no light ray 
comes out of it.
Furthermore in this region the roles of the time and radius are 
exchanged, i.e.~the curves $r=const.$ are space-like geodesics there.
This means that {\it all null and time-like geodesics with $r<r_H$ will reach 
the singularity at $r=0$}.
The nature of this singularity is analysed below.
The boundary of the trapped region is the {\it apparent horizon}
located at $r=r_H$ if $x^+>x^+_0$.

The line element (\ref{dsrt}) is not Minkowskian far from the black hole.
It is useful to introduce in this connection the new set of coordinates 
$(T,R)$ defined by $T=t$ and
\begin{eqnarray}
R+R_H\log\left\vert\,R-R_H\,\right\vert =
r_H \log\left\vert\,r-r_H\,\right\vert.
\label{R}
\end{eqnarray}
By assuming that the origins of the coordinates $r$ and $R$ coincide,
one deduces $R_H=r_H$.
Inside the collapsing shell the line element is not Minkowskian in these new
coordinates, but outside of it and very far from the black hole it is
asymptotically Minkowskian:  
\begin{eqnarray}
ds^2 = {\rm sgn}\left(R-\frac{1}{2M}\right)\, 
\left[\,dT^2- \frac{dR^2}{\left(\,1-\frac{1}{2MR}\,\right)^2}\,\right].
\label{dsRT}
\end{eqnarray}
This line element exhibits a singularity in the limit $R\rightarrow 0$, 
which does not appear in the metrics (\ref{dsx}), (\ref{dsy}) and
(\ref{dsrt}).
So one may suspect that this is due to the bad behaviour of the coordinate $R$ 
and that this is not a property of space-time itself.
But this is not the case.
That the central singularity is {\it real} comes from the fact that it
takes a {\it finite} amount of proper time or affine parameter for all 
time-like or null geodesics inside the apparent horizon to fall into 
$r=0$, and that it does not make any sense to extend them further \cite{HEWJ}.
One expects indeed a space-time singularity to be located at $r=0$ since 
all the energy is concentrated there after the collapse. 
This singularity has the topology of a {\it corner} because the scalar 
curvature vanishes along any curve ending at $r=0$, but is not defined there.
The remarkable result is that, contrary to the 3-D case \cite{SGA}, but like
the 4-D case \cite{Sy}, {\it the central singularity is located inside an 
apparent horizon}.
The line element (\ref{dsRT}) describes thus a black hole.
The constant $M$ is identified with its mass because of Eq.~(\ref{EM})
\cite{tH}.


The existence of a spontaneous creation of particles in this space-time is 
shown by calculating the expectation value of the energy-momentum tensor in 
the incoming vacuum.
This quantity will be denoted by $T_{\mu\nu}^{\rm rad}$.
Outside the imploding shell and far from the black hole, Eqs.~(\ref{y}), 
(\ref{rt}) and (\ref{R}) imply that $x^\pm\approx x^\pm(Z^\pm)$, 
where $Z^\pm=T\pm R$.
For both the scalar and Dirac massless fields, $T_{\mu\nu}^{\rm rad}$ is 
thus given in this region in terms of the Schwartzian derivative of 
$x^\pm(Z^\pm)$ \cite{Ve,DFU},
\begin{eqnarray}
T_{\pm\pm}^{\rm rad}\left(Z^\pm\right) \approx -\frac{1}{24\pi}\,
\left[\,\partial^2_\pm \ln\partial_\pm x^\pm -
\frac{1}{2} \left(\partial_\pm \ln\partial_\pm x^\pm\right)^2\,\right],
\nonumber
\end{eqnarray}
and $T_{+-}^{\rm rad} \approx 0$.
Noting furthermore that $x^+(Z^+)\approx Z^+$ and 
$x^-(Z^-)\approx x^-_H-\exp(-MZ^-)$, one obtains in this region
\begin{eqnarray}
\begin{array}{rlccrlc} 
\displaystyle T_{++}^{\rm rad}\left(Z^+\right) &\approx& 0, &\hspace{5mm}&
\displaystyle T_{--}^{\rm rad}\left(Z^-\right) &\approx& 
\displaystyle \frac{M^2}{48\pi}\cdot
\end{array}
\end{eqnarray}
This shows that the black hole emits a radiation.
Since the thermal average of the energy-momentum tensor equals 
$\frac{\pi k^2}{12\hbar^2}\,\Theta^2$ \cite{Ve}, where $\Theta$ is the 
temperature, then
\begin{eqnarray}
\Theta^{\rm rad} = \frac{\hbar M}{2\pi k}
\end{eqnarray}
is the temperature of the outgoing radiation detected by a distant 
{\it inertial} observer outside the imploding shell.

The back-reaction on space-time of the emitted radiation is analysed 
to one-loop order by adding to Eq.~(\ref{RT}) the trace anomaly 
$T^{\rm rad}(x)=-\frac{\hbar}{24\pi}R(x)$ \cite{DFU,BF}.
One obtains thus
\begin{eqnarray}
R(x) = \alpha\,8\pi G \, T^{\rm cl}(x)
\end{eqnarray}
where $\alpha=\frac{1}{1+\hbar G/3} < 1$.
The mass and temperature of the black hole are then renormalised by a term of 
order Planck length squared and become smaller:
\begin{eqnarray}
\begin{array}{rclcrcl}
M_{\rm BR} &=& \alpha\,M,  &\hspace{5mm}&
\Theta^{\rm rad}_{\rm BR} &=& \displaystyle 
\alpha\,\frac{\hbar M}{2\pi k}\cdot
\end{array}
\end{eqnarray}
Furthermore from Eq.~(\ref{rH}) its radius becomes larger.


If the ``$R=T$'' theory is considered instead of the 2-D theory of general 
relativity, the results of the present Letter show that there is a
2-D black hole whose metric, in a given set of coordinates, is Minkowskian 
far away.
This 2-D black hole differs in essence from the Schwarzschild black hole by
the fact that its central singularity is not {\it physically genuine} 
although real \cite{HEWJ}, because the curvature does not diverge in its 
neighbourhood.
Also, contrary to the 4-D case, this 2-D black hole grows without limits when 
its mass vanishes.
Although the line element (\ref{dsRT}) may not be directly obtained as a 
static solution from the equation of motion, it is physically relevant because
it describes the result of a collapse. 


I would like to thank C.~Isham, M.~E.~Ortiz, and S.~Schreckenberg
for helpful conversations and critical reading of the manuscript.
I also acknowledge support from the Soci\'et\'e Acad\'emique Vaudoise and 
from the Swiss National Science Foundation.


\begin{references}
\bibitem{DF} If the delta function had been defined by
$\int_{-\infty}^{+\infty}du \, f(u)\,\delta(u)=f(0^-)$, 
then Eqs.~(\ref{C(x)}) and (\ref{BC}) would imply
$\partial_- \ln \vert C(x)\vert = m(x^+)$ and
$C(x)= \exp\left[m(x^+)\,(x^--\Delta)\right]$,
where $\Delta$ is a constant and $m(x^+)$ is defined as above.
The problem is not well defined if
$\int_{-\infty}^{+\infty}du \, f(u)\,\delta(u) =f(0)$.
\end{references}
\end{document}